\documentclass[10pt, conference, compsocconf, a4paper]{IEEEtran}
\pdfoutput=1
\usepackage{ae} 
\usepackage{amsthm}
\usepackage[cmex10]{amsmath}
\usepackage{amssymb}
\usepackage{mathabx}
\usepackage[pdftex]{graphicx}
\usepackage{flushend} 
\usepackage{array}
\usepackage{url}

\usepackage{algorithmic}
\usepackage{algorithm}

\begin{document}

\title{{Performance of the Fuzzy Vault for Multiple Fingerprints}\\{\Large (Extended Version)}}

\author{\IEEEauthorblockN{Johannes Merkle, Matthias Niesing, Michael Schwaiger}
\IEEEauthorblockA{secunet Security Networks AG\\D-45128 Essen, Germany\\ 
johannes.merkle@secunet.com, \\ matthias.niesing@secunet.com\\ michael.schwaiger@secunet.com}
\and
\IEEEauthorblockN{Heinrich Ihmor, Ulrike Korte}
\IEEEauthorblockA{Bundesamt f\"ur Sicherheit in der Informationstechnik\\
D-53175 Bonn, Germany\\
heinrich.ihmor@bsi.bund.de\\ ulrike.korte@bsi.bund.de
}
}

\maketitle

\begin{abstract}
The fuzzy vault is an error tolerant authentication method that ensures the privacy of the stored reference data.  
Several publications have proposed the application of the fuzzy vault to fingerprints, but the results of subsequent analyses 
indicate that a single finger does not contain sufficient information for a secure implementation.   

In this contribution, we present an implementation of a fuzzy vault based on minutiae information in several fingerprints 
aiming at a security level comparable to current cryptographic applications. We analyze and empirically evaluate the security, 
efficiency, and robustness of the construction and several optimizations. The results allow an assessment of the capacity of the scheme and 
an appropriate selection of parameters. Finally, we report on a practical simulation conducted with ten users. 

This article represents a full version of \cite{MNS10c}, copyright of Gesellschaft f\"ur Informatik (GI). 
\end{abstract}

\section{Introduction}
Biometric authentication requires the storage of reference data for identity verification, either centrally (e.g., in a database) or locally 
(e.g., on the users token). However, the storage of biometric reference data poses considerable information security risks to the biometric application 
and concerns regarding data protection.
As a potential solution to this dilemma, {\it biometric template protection} systems \cite{BBG08} use reference data which reveal only very limited
information on the biometric trait. Another term frequently used for these schemes is {\it biometric encryption}. 
One of the most prominent approaches is the fuzzy vault scheme \cite{JS02}, where the sensitive information (the 
biometric reference data) is hidden among random {\it chaff points}. 

In \cite{CKL03,UPJ05,UJ06,NAP06}, the application of the fuzzy vault scheme to minutiae ({\it fuzzy fingerprint vault}) has been 
proposed. However, a subsequent analysis in \cite{M07} has revealed that the parameters suggested do not provide security beyond 50 bit cryptographic 
keys. One of the suggestions in \cite{M07} was to enhance security by using multiple fingers. This idea is supported by the observation in 
\cite{P09} that a single fingerprint cannot provide enough entropy to implement a secure biometric template protection. Further motivation for 
extending the fuzzy fingerprint vault to multiple fingers is provided by the analysis carried out 
in \cite{MIKNS10}, which shows that the number of minutiae needed to obtain a provable secure scheme based on the results of \cite{DORS08} is much larger than the number of minutiae typically detected on a single fingerprint.  
 
In this paper we present an implementation of a fuzzy vault based on the minutiae data of several fingerprints. We investigate the security,  
efficiency, and robustness of the scheme and of several optimizations applied, 
some of which have already been proposed in the previous constructions \cite{CKL03,UPJ05,UJ06,NAP06}. In particular, we 
evaluate the impact of the basic parameters and optimizations to error rates, efficiency and security, and we derive suggestions for 
parameter selection. Furthermore, we evaluate the practical performance of the scheme in an experiment with 10 users.

This article is structured as follows: In Section \ref{scheme}, we specify the fuzzy multi-fingerprint vault and its optimizations 
and justify our design decisions. 
Section \ref{analysis} recapitulates existing security results and contains additional security analysis of optimizations applied. 
In Section \ref{results}, we report the results obtained in evaluation with real fingerprints. Finally, in Section \ref{conclusions}, 
we draw conclusions and identify 
open issues for future investigations.

\section{Design of the scheme}\label{scheme}

\subsection{Basic biometric feature}\label{features}
We base our  biometric feature vectors on minutiae. Since the minutiae orientations 
resemble the orientation of the ridges at the minutiae position, they bear strong dependencies with the minutia location and with the orientation of other minutiae. Although such dependencies are hard to quantify, they could be exploited by sophisticated attacks to effectively reduce the search space. 
Furthermore, in the fuzzy vault 
scheme the stored minutiae are hidden in a large set of 
chaff points which must be indistinguishable from real minutiae; this objective is much 
harder to achieve if minutiae orientations are used as well. For these reasons we compose the feature vector of the minutiae 
location information (and the indexes of the corresponding fingers) only.   


\subsection{Underlying biometric template protection scheme}\label{biocrypt}
In the fuzzy vault scheme \cite{JS02}, a polynomial is used to redundantly encode a set of (pairwise distinct) private attributes $m_1, \ldots, m_t$ (e.g., biometric 
feature data) using a variant of Reed-Solomon decoding. First, a random (secret) polynomial
 $P(z)$ over a finite field $\mathbf{F}_q$ with degree smaller than $k$ is chosen. 
Then, each attribute $m_i$ is encoded as element $x_i$ of the finite 
field, i.e. $x_i=E(m_i)$, where $E$ is an arbitrary injective map from the space of attributes to $\mathbf{F}_q$. Each of these elements $x_i$ is evaluated over the polynomial, resulting in a list of (pairwise distinct) pairs $(x_i, y_i)\in \mathbf{F}_q^2$ with $y_i = P(x_i)$. 
In order to hide the private attributes, $r-t$ {\it chaff points} $x_{t+1}, \ldots, x_r \in \mathbf{F}_q$ are randomly selected so that $x_i \neq x_j$ for all $1\leq i < j \leq r$. For each chaff point $x_i$, a random $y_i\in \mathbf{F}_q$ with $y_i \neq P(x_i)$ is chosen. The list of all pairs $(x_{1},y_{1}), \ldots, (x_r,y_r)$, sorted in a predetermined order to conceal 
which points are genuine and which are the chaff points, is stored as the {\it vault}. 

For authentication and recovery of the secret polynomial, another set of attributes (the query set) has to be presented. This set is compared with the 
stored fuzzy vault $(x_{1},y_{1}), \ldots, (x_r,y_r)$, and those pairs $(x_i,y_i)$ are selected for which $x_i$ corresponds to an attribute 
in the query set. The selected points are then used to try to recover the secret polynomial using Reed-Solomon decoding. 

If the number of genuine points among the identified correspondences ({\it correct matches}) is at least $k$, 
the secret polynomial can be recovered. However, if the set of correspondences also comprises chaff points ({\it false matches}), 
the number of correct matches must be greater than $k$, or the decoding must operate on subsets of the matches resulting in many trials. Details are 
given in Section \ref{recovery}.

In the original fuzzy vault scheme, correspondence between points in the query set and the fuzzy vault means equality (of the encodings in the 
finite field). However, for the application of the 
fuzzy vault to fingerprints the definition of this correspondence is usually adjusted to allow a compensation of noise in the measurement of the minutiae. Following the approach of \cite{NAP06} and \cite{UJ06}, we define correspondence as mappings 
determined by a minutiae matching algorithm (see Section \ref{matcher}). 

The fuzzy vault scheme is error tolerant with respect to the set difference metric, which covers exactly the errors introduced to 
(naively encoded) minutiae information by insertions, omissions, and permutation of minutiae. The deployment of a minutiae matching algorithm 
for identifying correspondences between the query set and the fuzzy vault adds robustness with respect to global rotations and translations or non-linear deformations of the fingerprint. 
Since the matching algorithms included in standard fingerprint software only outputs a match score and not the 
list of corresponding minutiae, we use our own matching algorithm (see Section \ref{matcher}). 

In some cases even a random set of points from the fuzzy vault will result in the recovery of a polynomial, but with high probability it will 
not be the polynomial $P$ derived from the secret. In order to allow verification of the correctness of the recovered polynomial, we amend the 
scheme by storing a hash value of the polynomial's coefficients. If a secure cryptographic hash function is used, we can assume that the hash value
can not be used to recover the secret polynomial.

\subsection{Multi-biometric fusion}\label{fusion}
In order to obtain sufficient information for a secure scheme, we use multiple instances of fingerprints. 
Specifically, we use the imprints of $f\geq 2$ fingers of each person. 
The information extracted from the individual instances (the fingers used) can be merged at various levels \cite{N08}: 
at sensor level, feature level, score level, and decision level. Sensor level fusion introduces additional complexity when 
single finger sensors are used, because they need to be merged after acquisition. On the other hand, score level fusion and 
decision level fusion are generally not eligible for multi-instance biometric template protection,
because it implies the usage of individual secrets (polynomials in the case of the fuzzy vault) for each feature or instance; this 
enables separate brute force 
attacks on each biometric instance (in our case on each finger) individually, resulting only in a linear increase of security with the 
number of instances (fingers) used.
In contrast, in case of feature level fusion a single secret key is bound to the merged information of all instances, 
and a brute force attack needs to determine the biometric information of all instances 
simultaneously, resulting in an exponential increase of security with the number of instances.

For these reasons, we implemented feature level fusion by encoding the minutiae of all fingers in one feature vector. In this vector 
each minutiae is encoded as a triplet $(\theta,a,b)$, where $\theta\in \{1,\ldots, f\}$ is an index of the finger on which the minutiae 
was detected, while $a$ and $b$ denote the Cartesian coordinates of the minutiae location in the fingerprint image of the respective finger. Chaff points are encoded analogously. 

\subsection{Minutiae matching algorithm}\label{matcher}
We need to identify matching minutiae between fingerprints for enrollment and for verification: during enrollment minutiae matching is used to 
identify the most reliable minutiae from several measurements. During verification we have to identify a sufficiently large set of genuine 
minutiae within the vault to recover the secret polynomial. 

The matching is performed for each finger separately by a simple matching algorithm that 
identifies minutiae correspondences between two sets $M$ and $\bar{M}$ of points $\mathbf{m}$ 
(minutiae or chaff points) which are given by their positions $(a, b)$  
in the fingerprint image.
The algorithm tries to maximize the number of correspondent points between the sets by finding a suitable 
global rotation and translation transformation $T$ and tolerates (Euclidean) distances $||\cdot ||_2$ 
between two points smaller than $\delta$, where $\delta$ is a parameter of the algorithm. Further parameters are a tolerance $\epsilon$ for the comparison 
of distances between minutiae and the limits for rotation $\omega$ and translation $S$, which are introduced to limit the computational complexity of 
the algorithm. 

\begin{algorithm}[htb]
\caption{Minutiae Matching Algorithm}
\label{matchalg}
\begin{algorithmic}
\STATE {\bf Input:} Sets $M=\{\mathbf{m}_1,\ldots,\mathbf{m}_n\}$ and $\bar{M}=\{\mathbf{\bar{m}}_1,\ldots,\mathbf{\bar{m}}_{n'})\}$ of 
(minutiae or chaff) points with $\mathbf{m}_i, \mathbf{\bar{m}}_i \in \mathcal{E}$. \smallskip

\FORALL{$1 \leq i < j \leq n$} 
\STATE Set $D_{ij} = ||\mathbf{m}_i-\mathbf{m}_j||_2 $.
\ENDFOR  

\FORALL{$1 \leq i' < j' \leq n'$} 
\STATE Set $D'_{i'j'} = ||\mathbf{\bar{m}}_{i'}-\mathbf{\bar{m}}_{j'}||_2$.
\ENDFOR  

\FORALL{$1 \leq i < j \leq n$}
\FORALL{$i',j'$ with $|D_{ij}-D_{i'j'}| \leq \epsilon$}
\STATE Identify the isometry $T=(\phi, \mathbf{v})$ so that $T(\mathbf{m}_i)=\mathbf{\bar{m}}_{i'}$ and 
$T(\mathbf{m}_j)=\mathbf{\bar{m}}_{j'}$.
\IF{rotation and translation are within the configured limits, i.e., $|\phi| < \omega$ and $\| \mathbf{v}\|_2 < S$}  
\STATE Apply isometry $T$ to all points in B.  
\STATE Identify set $C_{T}$ of pairs $(\mathbf{m}_{\alpha},T(\mathbf{\bar{m}}_{\beta}))$ so that 
$\mathbf{m}_{\alpha}$ and $T(\mathbf{\bar{m}}_{\beta})$ have distance smaller than $\delta$ and are the closest matches for each other. 
\ENDIF
\ENDFOR  
\ENDFOR  
\RETURN Isometry $T$ and set of minutiae mappings $C_{T}$ for which $|C_{T}|$ is maximal.
\end{algorithmic}
\end{algorithm}

A pseudo code description is given in Algorithm \ref{matchalg}. 
Since we apply the minutiae matching algorithm for each finger separately, no finger index is used. 

Our experiments revealed that $\epsilon=0.2$, $\omega=45^{\circ}$, and $S=200$ (pixels) already yield relatively good matching results (after 
pre-alignment of the images as described in Section \ref{prealignment}), while larger values increase the running time considerably without significant 
improvement of the matching rate. Therefore, we fixed these parameters to these values. 

The tolerance parameter $\delta$ varies: we use a greater value $\delta=\delta_{\mathrm{e}}$ for enrollment than the value 
$\delta=\delta_{\mathrm{V}}$ for verification to increase the number of reliable minutiae.

\subsection{Optimizations}
\subsubsection{Restriction of fingerprint area}\label{ellipse} 
\begin{figure}[bt]                                  
\centering                                          
\includegraphics[width=0.3\textwidth]{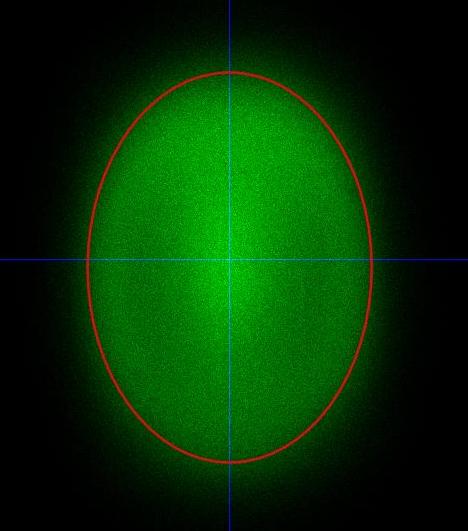}
\caption{Distribution of minutiae positions in 82800 fingerprints and the ellipse $\mathcal{E}$ from where minutiae are considered. 
The brightness of pixels corresponds to the frequency of minutiae occurrence at this position.}            
\label{fig:mindistr}
\end{figure}

Since fingerprints usually assume an oval shape, minutiae rarely occur in the corners of the image, provided that the sensor area is sufficiently large. 
In order to ensure that the distribution of the randomly selected chaff points resembles that of genuine minutiae, we restrict both the chaff points 
and the minutiae considered in the vault to an area $\mathcal{M}$ with sufficiently high minutiae occurrence. Such an area was empirically 
determined by a statistical evaluation of the positions of 5.8 million minutiae extracted from 82800 imprints taken from 9200 fingers with 3 
different sensors having 500 DPI.\footnote{The fingerprints had been collected in a previous project of the BSI and secunet.} It turned out that - 
independent of the finger - $7/8$ of all minutiae occurred in an 
area defined by an ellipse $\mathcal{E}$ that covers approximately 87000 pixels, which roughly corresponds to $2.25\;\mathrm{cm}^2$. 
The distribution of the minutiae positions and the ellipse are shown in Figure \ref{fig:mindistr}. 
Consequently, we choose minutiae and chaff points only from the set $\mathcal{M}$ given by the union of these ellipses on the considered fingers, i.e. $\mathcal{M}:=\{(\theta,a,b) \, | \, \theta \leq f \wedge (a,b) \in \mathcal{E} \}$.

\subsubsection{Reliability filtering during enrollment}\label{filter} 
In order to minimize minutiae insertion and omission errors, we use only the most reliable minutiae 
for the feature vector. Several minutiae quality assessment indices have been developed (e.g.\ \cite{WGT07} or \cite{CDJ05}) which could be 
used to (\`{a} priori) predict the reliability of detection of the individual minutiae. However, we anticipate that an empirical 
(\`{a} posteriori) determination of the detection reliability with a sufficient number of samples yields better results. 
For this reason, we use multiple measurements during enrollment and consider only those minutiae in the feature vector that have been detected in all measurements. Details are given in Section \ref{enrollment}.

\subsubsection{Enforcing minimum distance}\label{mindistance}
Due to the deviations in the measured minutiae locations, it can happen during authentication that a minutia in a query fingerprint 
is closer to a chaff point than to the corresponding minutiae in the vault. Although we try to minimize these deviations by setting 
the locations of the stored minutiae to their mean value measured during enrollment, the frequency of such assignment errors can drastically 
increase if the chaff points are selected too close to genuine minutiae. Therefore, we select the chaff points with a minimum distance $d$ to the genuine minutiae of the same finger with respect to the Euclidean distance. Furthermore, in order to prevent that an adversary can exploit this minimum distance to distinguish chaff points 
from genuine minutiae, we also enforce the minimum distance among the minutiae and chaff points. In particular, 
if the Euclidean distance between two minutiae on the same finger is smaller than $d$, one of them is randomly disregarded, and chaff points are selected with minimum Euclidean distance $d$ to all other chaff points and minutiae from the same finger.

\subsubsection{Quality filtering during authentication}\label{quality_filter}
In \cite{MIKNS10}, it is shown that the achievable entropy loss of the scheme decreases with an increase of the average number of surplus minutiae 
(i.e., minutiae not matching with real minutiae in the reference template) per query fingerprint. The reason for this fact is that, on 
average, the number of {\em false matches} of minutiae with chaff points increases with the average number of surplus minutiae per query fingerprint. However, an increase of 
false matches requires stronger error correction by lowering the degree $k$ of the secret polynomial, which decreases security of the scheme with respect to 
both information theoretic lower bounds and practical attacks. 

In order to limit the average number $s$ of surplus minutiae per query fingerprint, we filter the minutiae from the query fingerprint using the quality index value output by the 
minutiae extraction algorithm. Precisely, we define a minimum quality value $Q$ and provide to the matching algorithm only those minutiae of the 
query fingerprint that have a quality value of at least $Q$. In our concrete implementation we used the MINDTCT algorithm of NIST \cite{WGT07} for 
minutiae extraction which outputs minutiae quality values in the range between 0 and 1.  

The motivation for the quality filtering during authentication is that during enrollment 
only the most reliable minutiae have been used for computing the reference template. Since the quality values of the minutiae should predict 
their detection reliability, minutiae with higher quality value are more likely to be used for template generation. In practice, however, the quality 
value of a minutiae considerably varies between different measurements. Furthermore, experiments show that reliable minutiae sometimes have small 
quality values. This implies that quality filtering would not only reduce the number of false matches but also the number of correct matches. 
Therefore, the extent of filtering, i.e., the value chosen for parameter $Q$, must be carefully selected based on empirical data so that the number of
correct matches is not significantly reduced.

\subsubsection{Enforcement of minimum number of minutiae per finger}\label{chi}
One of the main sources for failures during authentication is the difficulty to correctly align the query fingerprints 
with respect to the stored minutiae. This task is performed by the minutiae matching algorithm (described in Section \ref{matcher}) 
for each finger by identifying the isometry (rotation and translation) that maximizes the number of matches between the minutiae extracted 
from the query fingerprint and the points (representing minutiae or chaff point) stored in the reference data. However, this approach can only 
be successful if the reference template contains a sufficient number of minutiae of each finger; otherwise, i.e., if for 
one of the fingers the reference template contains only few minutiae, the number of wrong matches (with 
chaff points) resulting by chance from an incorrect isometry may be higher than the number of matches for the correct isometry. 
In practice, such cases can easily occur if one of the fingerprints captured during enrollment is of relatively poor quality.
 
For this reason, we require that the reference template computed during authentication contains at least $\chi$ minutiae from each finger, where 
$\chi$ is an additional parameter. Since this constraint reduces the number of possible reference templates its impact on security must be 
analyzed. We provide an estimation of this reduction in Section \ref{analysis_chi}.  

\subsubsection{Pre-alignment of fingers and threshold for rotation}\label{prealignment}
Multi-biometric systems are more laborious and time-consuming for the user than single-biometric systems. This is particularly true for 
the fuzzy fingerprint vault: for security  reasons (see Section \ref{fusion} for a discussion), the success of the authentication is not evaluated for 
each finger individually but only for all of them together; this implies that in case of a failure the user does not know which of the fingerprints 
caused the authentication to fail and, hence, he is forced to re-capture all fingers. However, the following discussion shows that poor 
quality fingerprints can be quite reliably detected even before the polynomial reconstruction is started.  

Our matching algorithm does not rely on a correct alignment of the query fingerprint in relation to the minutiae set contained in the 
reference template, but it identifies an isometry (rotation and translation) between the set of minutiae of the query fingerprint and the points of 
the reference template so that the number of matches is maximized. In most cases this approach works well and the matching algorithm outputs 
many more correct matches than false matches (with chaff points) which indicates that the isometry identified represents the correct alignment 
of the fingerprints quite exactly. However, in some few cases the number of wrong matches is greater than the number of correct matches, which 
typically results in a failure to recover the secret polynomial. Our analysis of more than 160 tests performed shows that in most of these cases
the rotation applied by the matching algorithm was extraordinarily large, indicating that the identified isometry was incorrect. 

This observation shows that a threshold for the rotation identified by the matching algorithm could be used to detect poor quality fingerprints 
that result in failed verifications. However, in order to maximize the efficiency of this approach, we should try to minimize the magnitude of 
the rotation in the correct isometry, i.e., the isometry that correctly maps the minutiae of the query fingerprint to the minutiae in the 
reference template. We do so by pre-aligning the fingerprints before the minutiae are extracted. Several approaches have been proposed for this 
task, in particular the detection of singular points (e.g., \cite{LJK05}) or the additional storage of supplementing alignment data (e.g., \cite{JA07, LYTSL08}). However, alignment data may reveal information about the minutiae and singular point detection is quite complex. Therefore, we 
decided to follow a different approach which is simple and does not require the storage of additional reference data. 

\begin{figure}[bt]                                  
\centering                                          
\includegraphics[width=0.3\columnwidth]{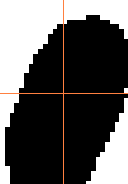}
\caption{Uneven distribution of the pixels between the pairs of opposed quadrants triggering a rotation counterclockwise.}            
\label{fig:rot}
\end{figure}

Our pre-alignment algorithm scales down the fingerprint image and uses a threshold on pixel brightness to obtain an image displaying the shape of the fingerprint as black area. Furthermore, the image is shifted so that the centroid of all black pixels matches the image center. 
Then it iteratively performs the following step: 

If the sum of black pixels in the upper left and the lower right quadrant 
exceeds the number of black pixels in the lower left and the upper right quadrant, the image is rotated clockwise by $1^{\circ}$; else, it is rotated 
counterclockwise by $1^{\circ}$. Following each rotation the wedge at the lower end resulting from the rotation is removed by horizontal cropping.  
The evaluation criteria for the rotation is illustrated in Figure \ref{fig:rot}.

The iteration stops as soon as the direction of rotation changes. Finally, the 
aggregated rotations are applied to the original fingerprint image.

\subsection{Enrollment}\label{enrollment}
Let $f\geq 2$ be the number of fingers used per person, $q$ a prime power, $k < t < r \leq q$, and $\chi \leq t/f$. 
For each user, a random polynomial $P$ of degree less than $k$ over the finite field $\mathbf{F}_q$ is selected. 
The coefficients of this polynomial represent the secret of the scheme. Then, for each finger $u$ imprints 
are taken and the minutiae correspondences between these instances are identified using the matching algorithm (Algorithm \ref{matchalg}) 
with tolerance parameter $\delta=\delta_{\mathrm{e}}$. Minutiae outside the considered set $\mathcal{M}$, i.e., with position outside the 
ellipse $\mathcal{E}$ on the respective finger, are neglected (see Section \ref{ellipse}). 
Then, $t$ of those minutiae that have been detected in all $u$ imprints of the respective 
finger are selected at random so that at least $\chi$ minutiae are taken from each finger and each pair of chosen minutiae has a minimum distance 
of $d$. 
This set $T$ of $t$ reliable minutiae can be considered the biometric template to be protected by the fuzzy vault scheme.
The template $T$ is amended with random chaff points,
 resulting in a set $R$ of $r$ points 
containing $t$ genuine minutiae and $r-t$ chaff points, so that each point in $R$ has a minimum distance of $d$ to all other points from the same finger. 
Furthermore, in order to ensure that minutiae and chaff points within 
the vault are not distinguishable by their order, they are lexicographically ordered. 

In contrast to the original fuzzy vault scheme \cite{JS02}, the secret polynomial is redundantly encoded not by evaluating it on the biometric 
data itself but only on the minutiae's indexes in the ordered list. Precisely, for all 
$1\leq i \leq r$ we (re-)define $x_i=E(i)$, where $E$ is an injective embedding from the set $\{1,\ldots,r\}\subset \mathbf{Z}$ to $\mathbf{F}_q$. (For instance, if $q$ is prime we could set $E(i):=i \bmod q$.) Further, we set 
$y_i=P(x_j)$, if $\mathbf{m}_{i}$ is a genuine minutia, and choose a random value $y_i \neq P(x_j)$, if $\mathbf{m}_{i}$ is a chaff point, where $j$ is the index of $\mathbf{m}_{i}$ after applying the lexicographic order. This optimization allows a reduction of the field size to the range of $r$. 
The {\em vault} $Y$ is given by  
the ordered list of minutiae and chaff points, paired with the corresponding $y_j$ values. The vault and a hash value $H$ of the polynomial's 
coefficients are stored in the database.

\begin{algorithm}[htb]
\caption{Enrollment}\label{alg_enrol}
\begin{algorithmic}
\STATE Select a random polynomial $P(z)=\sum_{i=0}^{k-1}e_i z^{i}$ over $\mathbf{F}_q$. 
\FOR{$\theta = 1$ to $f$} 
\STATE Take $u$ imprints of finger $\theta$ and pre-align them (see Section \ref{prealignment}) 
\STATE Extract the corresponding sets $M_{\theta,1},\ldots,M_{\theta,u}$ of minutiae.
\FOR{$j=2$ to $u$}
\STATE Find correspondences between $M_{\theta,1}$ and $M_{\theta,j}$ using Algorithm \ref{matchalg} with $\delta=\delta_{\mathrm{e}}$. 
\ENDFOR
\STATE Identify the set $A_\theta$ of {\it reliable minutiae} that are present in all $u$ imprints. 
\FORALL{Minutiae $\mathbf{m}\in A_\theta$} 
\STATE Set $\mathbf{m}=(\theta,a,b)$, where $(a,b)$ is the mean value of its location over the $u$ measurements.
\ENDFOR
\STATE Remove all $(\theta,a,b)$ from $A_\theta$ with $(a,b)\not \in \mathcal{E}$.
\WHILE{two minutiae have distance smaller than $d$}
\STATE Randomly remove one of them.
\ENDWHILE
\IF{$|A_\theta|<\chi$}
\STATE Repeat capture process for finger $\theta$.
\ENDIF
\ENDFOR
\STATE Set $A=\bigcup_{\theta} A_\theta$ as the set of all reliable minutiae.
\IF{$|A| < t$}
\RETURN{``ERROR: Not enough reliable minutiae''}
\ENDIF
\STATE Randomly select $T=\{\mathbf{m}_1,\ldots,\mathbf{m}_t\} \subseteq A$ so that $T$ contains at least $\chi$ minutiae for each finger. 
\STATE Randomly select $r-t$ chaff points $\mathbf{m}_{t+1},\ldots,\mathbf{m}_r$ so that $||\mathbf{m}_i-\mathbf{m}_j||\geq d$ for all 
$1 \leq i < j \leq r$.
\STATE Compute the lexicographic order $(\mathbf{m}_{j_1},\ldots,\mathbf{m}_{j_r})$ of $R=\{\mathbf{m}_1,\ldots,\mathbf{m}_r\}$.
\FOR{$i=1$ to $r$}
\IF{$\mathbf{m}_{j_i}$ is a genuine minutia, i.e., $j_i \leq t$}
\STATE Set $y_{i}=P(x_i)$ with $x_i=E(i)$.
\ELSE 
\STATE Select a random $y_{i}\neq P(x_i)\in\mathbf{F}_q$.
\ENDIF
\ENDFOR
\STATE Set $Y=(\mathbf{m}_{j_1},y_{1},\ldots,\mathbf{m}_{j_r},y_{r})$.
\STATE Compute $H:=h(e_0|| \ldots ||e_{k-1})$, where $||$ denotes concatenation.
\STATE Store $Y$ and $H$ in database. \end{algorithmic}
\end{algorithm}

A pseudo code description of the enrollment is given in Algorithm \ref{alg_enrol}. There $h$ denotes a collision resistant hash function. 
Furthermore, both 
minutiae and chaff points are denoted as $\mathbf{m}=(\theta,a,b)$, where $\theta$ is an index of the finger and $(a,b)$ are the Cartesian coordinates of 
its location in the image. We define the {\it distance} $||\mathbf{m}-\mathbf{m'}||$ of two minutiae on the same finger as the Euclidean 
distance of their location.

\subsection{Recovery of the polynomial}\label{recovery}
The unlocking of the vault (during authentication) requires the recovery of the secret polynomial $P$ from a set 
of points $(x_{j_i},y_{j_i})$, some of which (those resulting from {\it correct matches} with minutiae) lie on the polynomial, while others 
(resulting from {\it false matches} with chaff points) do not. For this task, a Reed-Solomon decoder {\sc RSDecode} is used that receives as input
a set of $w$ points $(x_{j_1},y_{j_1}), \ldots, (x_{j_w},y_{j_w})\in \mathbf{F}_q^2$ with $w \geq k$ and outputs 
$e_0, \ldots, e_{k-1} \in \{0,\ldots, q-1\}$, so that $y_{j_i}=P(x_{j_i})$ holds for at least $k$ of the $(x_{j_i},y_{j_i})$ with 
$P(z)=\sum_{i=0}^{k-1}e_i z^{i}$, if such a polynomial exists. We assume that the Peterson-Berlekamp-Massey-decoder is used 
as suggested in \cite{JS02}. This technique is successful, if at least $(w+k)/2$ of the $w$ points handed over to the decoder 
are correct. Although there are Reed-Solomon-Decoders that can decode with only $\sqrt{wk}$ correct points, they do not offer any advantage 
for the fuzzy vault, because for typical parameters $\sqrt{wk}$ is quite close to $(w+k)/2$, and they are computationally much less efficient 
(see \cite{JS02}). 

As pointed out in \cite{CKL03}, the Reed Solomon Decoding degenerates to a brute-force polynomial interpolation for $w=k$. Previous 
implementations of the fuzzy fingerprint vault \cite{CKL03,UPJ05,UJ06} have used this brute-force approach for decoding. However, numerical evaluation reported in 
\cite{MIKNS10} revealed that setting $w=2 m_{\mathrm{c}} -k$ and $k\approx m_{\mathrm{c}}- m_{\mathrm{f}}$ can provide a good balance between efficient decoding and security, where $m_{\mathrm{c}}$ and $m_{\mathrm{f}}$ are the expected numbers of correct and false 
matches, respectively. However, if the match rate disperses considerably, it may by necessary to slightly deviate from this value, in order to reduce the False Rejection Rate. As we will see in Section \ref{match_rate}, this is the case.

\subsection{Authentication}
We only implement an authentication in the verification scenario, where the identity of the (alleged) user is known a priori. 

In order to verify the identity of a user, a query fingerprint is taken for each considered finger. The minutiae are extracted and matched 
with the minutiae and chaff points $\mathbf{m}$ contained in the vault stored for the alleged user. The indices of those minutiae 
and chaff points in the vault matching with minutiae in the query fingerprint are identified; along with the corresponding $y_i$ values 
they are given to {\sc RSDecode} (see Section \ref{recovery}) to recover the secret polynomial $P$. If the  number of genuine minutiae among
these points is sufficiently high (see Section \ref{recovery} for a discussion), the polynomial can be recovered. Finally, the correctness 
of the recovered polynomial is verified using the hash value stored in the database. Optionally, the secret key (given by the coefficients 
of the recovered polynomial) can be used for further cryptographic applications, e.g., as seed in a key derivation function. 

A pseudo code description of the verification is given as Algorithm \ref{alg_verify}. The tolerance parameter $\delta_{\mathrm{v}}$ used for 
the minutiae matching algorithm can differ from that used during enrollment. 

\begin{algorithm}[htb]
\caption{Verification of a user's identity}\label{alg_verify}
\begin{algorithmic}
\STATE Select $Y$ and $H$ stored in database for this user
\STATE Set $I=\emptyset$.
\FOR{$\theta=1$ to $f$} 
\STATE Identify in $Y$ the subset $R_\theta$ of minutiae and chaff points on finger $\theta$.
\STATE Take a query fingerprint for finger $\theta$ and pre-align it as described in Section \ref{prealignment}.
\STATE Extract the set $M_\theta$ of minutiae with quality value at least $Q$ (see Section \ref{quality_filter}).
\STATE Find correspondences between $R_\theta$ and $M_\theta$ using Algorithm \ref{matchalg} with $\delta=\delta_{\mathrm{v}}$.
\FORALL{$\mathbf{m}\in R_\theta$, which have matched a minutia in $M_\theta$} 
\STATE Add $(x_i,y_i)$ to $I$, where $x_i=E(i)$ and $i$ is the index of $\mathbf{m}$ in $Y$.
\ENDFOR
\ENDFOR
\FORALL{Subsets $Z$ of $I$ with cardinality $x$} 
\STATE Start {\sc RSDecode} on input $Z$ (see Section \ref{recovery}).
\IF{{\sc RSDecode} outputs $\bar{e}_0,\ldots, \bar{e}_{k-1}$.}
\IF{$\bar{H}=h(\bar{e}_0||\ldots || \bar{e}_{k-1})=H$}
\RETURN{``Verification successful''.}
\STATE Optional: Return $\bar{e}_0||\ldots || \bar{e}_{k-1}$.
\STATE Stop algorithm. 
\ENDIF
\ENDIF 
\ENDFOR
\RETURN{``Verification not successful''} 
\end{algorithmic}
\end{algorithm}

\section{Security analysis}\label{analysis}
In this contribution we consider the security of the fuzzy vault for multiple fingerprints 
with respect to attacks that try to recover the minutiae or, equivalently, the secret 
polynomial from the vault. It is understood that there are other types of attacks 
against biometric template protection schemes to which the fuzzy vault is susceptible 
\cite{sb07}. In particular, the cross matching of the vaults from several independent 
enrollments of a user represents a serious threat to the fuzzy vault. However, a comprehensive analysis of all potential attacks against the 
fuzzy vault would go beyond the scope of this paper.   

\subsection{Provable Security}\label{provable_security}
In \cite{MIKNS10}, lower bounds on the security of the fuzzy fingerprint vault were deducted. In particular, for the case $r=q$ and $\chi =0$, an upper bound was given for the success probability of any algorithm that takes as input the vault $Y$ and tries 
to output the corresponding secret polynomial or the corresponding template $T$.   
The analysis conducted in \cite{MIKNS10} revealed that a security of at least 50 bits can only be achieved 
if the match rate, i.e., the rate  
of minutiae in the vault matched with the query fingerprints during verification, exceeds a certain minimum value 
which depends on the average number $s$ of surplus minutiae  (i.e. the minutiae that do not match with a genuine minutia in the template) per query fingerprint and on the tolerance parameter $\delta_{\mathrm{v}}$ used for matching during verification. 

\subsection{Existing attacks}\label{practical}
The most efficient method to recover the minutiae or secret polynomial from the vault 
was published by Mihailescu \cite{M07}. This brute force attack is designed to break the implementations of \cite{CKL03}, \cite{UPJ05} 
and \cite{UJ06}; in the context of our scheme it is even slightly more efficient as the correctness of the recovered polynomial can be verified 
using the hash value of the secret coefficients and does not require additional evaluations of the polynomial. With this adaptation the attack 
systematically searches through all subsets $\{i_1,\ldots,i_k\}$ of $\{1,\ldots,r\}$, computes the unique polynomial $P$ satisfying
$P(E(i_j))=y_{i_j}$ by Lagrange interpolation, and checks the correctness of this polynomial with the 
stored hash value. According to \cite{M07}, the number of trials needed is $1.1 (r/t)^k$ and each trial requires $6.5 k \log^2{(k)}$ arithmetic 
operations over $\mathbf{F}_q$.\footnote{In fact, the number of trials needed is considerably higher, as the estimate $\binom{r}{k}/\binom{t}{k}< 1.1 
(r/t)^k$ used in \cite{M07} for $r>t>5$ does not hold true.} However, in the latter estimation an explicit constant of 18 for multiplication of the 
polynomials (see Corollary 8.19 in \cite{GG00}) has been overlooked, and thus, we end up with a total number of approximately $129 k \log^2{(k)} 
(r/t)^k$ arithmetic operations. 

If the number of chaff points is close to the maximum possible, the attack described in \cite{CST06} can be more efficient than brute force. The basic idea is that the free area around chaff points is smaller than around genuine minutia.  
Assuming a density 0.45 for random sphere packings \cite{CKL03}, the maximum number of chaff points per finger would be 
$0.45 \cdot 87000 / V_{d}$, where $V_d$ is the number of integer point in the sphere of radius $d$. 

Another threat to the fuzzy fingerprint vault is the {\em fingerprint dictionary attack}, where an attacker simulates verification with the vault using a large number of real or artificial fingerprints. If the templates are chosen according to a realistic distribution, the success probability of each attempt equals the False Accept Rate (FAR) and the attacker needs  $(\mathrm{FAR})^{-1}$ trials on average. Therefore, if a single simulation of  verification takes $C$ computational steps, the attack's running time is $C \cdot (\mathrm{FAR})^{-1}$. Note, that simulation involves templates extraction and minutiae matching, which are computationally much more expensive than a polynomial interpolation. 

Unfortunately, determination of very small FAR values is computationally very expensive: While the FAR for the multi-finger setting (i.e. for $f\geq 2$) can be extrapolated from the FAR of a single-finger setting, determination of latter one requires considerably more than $\mathrm{FAR}^{-1}$ matching operations and must be performed for each set of parameters individually. For the security level we aim at, this already takes more time than we have available for this publication. Therefore, we do not evaluate the security with respect to the fingerprint dictionary attack, but leave this task to a future publication.

\subsection{Entropy loss by the minimum number of minutiae per finger}\label{analysis_chi}
Whereas the enforcement of a minimum number $\chi$ of minutiae per finger (see Section \ref{chi}) 
aims at reducing the false rejection rate it also decreases the security of the scheme by 
narrowing the set of possible templates. This applies to the lower 
bound on attacks according to \cite{MIKNS10} as well as to the practical attack of \cite{M07}. The subsequent analysis quantifies this reduction of security.

We will assume that the minutiae chosen are independently and uniformly 
distributed among the $F$ fingers. This assumption can be fulfilled by a suitable 
probabilistic selection method of the template $T$ from the set of reliable minutiae during enrollment. 

Using this assumption and the inclusion-exclusion-principle, we can estimate the probability $\zeta(t,\chi)$ that a template with 
$t$ minutiae includes for each finger $f$ at least $\chi$ minutiae by
\begin{IEEEeqnarray*}{rCl}\label{eq:zeta}
\zeta(t,\chi) & = & 1 - f^{-t} \sum_{\theta=1}^{f}(-1)^{\theta} \binom{f}{\theta}  \\
& & \cdot \sum_{i_1,\ldots,i_\theta=0}^{\chi} \binom{t}{i_1, \ldots i_\theta, t-\sum_j i_j} 
(f - \theta)^{t-\sum_j i_j},
\end{IEEEeqnarray*}
where $\binom{a}{b_1, \ldots, b_m}$ with $b_1+ \cdots + b_m = a$ denotes the multinomial coefficient. 

On the other hand, the conditional probability $p$ that a particular instance of a template $T$ is chosen, if a minimum number of $\chi$ minutiae 
per finger is enforced, can be calculated from the probability $p'$ that this instance is chosen, if no minimum number of minutiae per finger is
enforced, by the equation $p= p'/ \zeta(t,\chi)$. Therefore, the entropy $\mathbf{H}_\infty(T_{\chi})$ of a template chosen with a minimum 
number of minutiae per finger is exactly $\log{\zeta(t,\chi)}$ smaller than the entropy of a template chosen without a minimum 
number of minutiae per finger. 

For practical attacks, the search space is narrowed by the factor $\zeta(t,\chi)$; thus, the best known attack could be adapted to require at most 
$129 \zeta(t,\chi) k \log^2{(k)} (r/t)^k$ operations.

\section{Results}\label{results}
In this section, we summarize the results of empirical parameter evaluations, the impact of the individual optimizations, and the
 general performance of the scheme. 
 
We used a test set of 864 fingerprints taken from 18 persons in the course of this research using an optical sensor, each person providing 6 
 imprints of 8 fingers (little fingers were excluded). In our experiments, we used 6 or all 8 fingers per person (without or with thumbs), but results referring to single fingers were averaged over all finger types.  

For minutiae extraction, we used the MINDTCT algorithm of NIST \cite{WGT07}. We stress that other feature extraction algorithms may exhibit a different performance, and therefore, the resulting statistics may deviate from ours. 
 
\subsection{Size of feature vector}\label{size}
First, we determined how large the feature vector can be in dependence of the number $u$ of measurements and the tolerance parameter 
$\delta_{\mathrm{e}}$ used during enrollment. We did this by evaluating the number of minutiae per finger that are reliably (i.e., $u$ times) 
detected in $u$ measurements. Since this number varies considerably among individuals and measurements, acceptable Failure To Enroll (FTE) rates can only be achieved, if the required number of reliable minutiae is considerably lower than its average value. Therefore, we evaluated the maximum number $M_{\mathrm r}$ of reliable minutiae that is achieved in at least 80\% of all measurements. The results of this evaluation are listed in Table \ref{tab:1}.

\begin{table}[tb]
\begin{center}
\caption{Number $M_{\mathrm r}$ of reliable minutiae per finger that is found in 80\% of all measurements.}
\label{tab:1}
\begin{tabular}{|c|cccc|}
\hline
u & $\delta_{\mathrm{e}}=5$ & $\delta_{\mathrm{e}}=7$ & $\delta_{\mathrm{e}}=10$ & $\delta_{\mathrm{e}}=15$ \\ \hline \hline
1 & 63 &	63 &	63 &	63 \\
2 & 23 & 32	& 39	& 43\\
3 & 18 & 24	& 31	& 35\\
4 &  9 & 16	& 22	& 27\\
5 &  6 &  9 & 15	& 18\\\hline\end{tabular}
\end{center}
\end{table}

In the multi-biometric setting, a single finger having only few reliable minutiae can be compensated by the others. However, in the case of only two 
fingers, this effect is smaller than for $f\geq 3$. Moreover, if an individual generally 
has low quality fingerprints, e.g. due to skin abrasion, cuts or dry skin, the probabilities that several fingers have few reliable minutiae are 
dependent and do not multiply. 
Therefore, the required number of reliable minutiae should be carefully selected based on empirical evaluation 
of the resulting enrollment error rates.

Due to the small number of reliable minutiae for $u=5$, we generally recommend to set $u \leq 4$. 

\subsection{Minutiae matching rates}\label{match_rate}
In order to configure the error correction capabilities of our scheme appropriately, it is necessary to determine the rate at which the genuine minutiae in the 
vault are identified during authentication. For various tolerance parameters $\delta_{\mathrm{v}}=\delta_{\mathrm{e}}$, 
we computed the biometric template set $T$, containing $t$ minutiae reliably detected in $u$ measurements, and matched them with the minutiae of an 
(independent) query fingerprint using our matching algorithm. We did not add chaff points to the template $T$. 
The average match rate, i.e. the average ratio between the number of matches found 
and $t$, are given in Table \ref{tab:2}.  

\begin{table}[tb]
\begin{center}
\caption{Average match rate (in percentage) in the absence of chaff points for $\delta_{\mathrm{e}}=\delta_{\mathrm{v}}$.}
\label{tab:2}
\begin{tabular}{|c|cccc|}
\hline
u & $\delta_{\mathrm{v}}=5$ & $\delta_{\mathrm{v}}=7$ & $\delta_{\mathrm{v}}=10$ & $\delta_{\mathrm{v}}=15$ \\ \hline \hline
1 & 40 &  50 &	58 &	64 \\
2 & 66 &	72 &	78 &	82 \\
3 & 75 &	81 &	84 &	87 \\
4 & 81 &	85 &	88 &	90 \\
5 & 85 &	89 &	91 &	92 \\ \hline
\end{tabular}
\end{center}
\end{table}

Similarly to the number of reliable minutiae, the match rate varies considerably between different measurements. Moreover, in the presence of chaff points, the match rates slightly decrease depending on the expected number of false matches (with chaff points),
as the chaff points render the correct mapping of the minutiae more difficult for the matching algorithm. (This aspect is further discussed in Section 
\ref{delta}.) Therefore, a reasonably small FRR can only be achieved if $k$ is selected slightly smaller than the expected value of $m_{\mathrm{c}}- m_{\mathrm{f}}$ (see Section \ref{recovery}). Our empirical evaluation suggests to set $k$ 10\%-20\% smaller than this value. 

For $2\leq u \leq 4$, we obtain good match rates at a reasonable number of minutiae. Therefore, we will subsequently focus on these cases. 



\subsection{Effect of quality filtering during verification}\label{res_qual}
As argued in Section \ref{quality_filter}, quality filtering of the minutiae in the query fingerprints aims to reduce the number of surplus minutiae, i.e., minutiae in the query fingerprints that do not match with genuine minutiae in $T$. We evaluated the effectiveness and eligible configuration of the filtering based on the minutiae quality values output by the MINDTCT algorithm of NIST \cite{WGT07}. 

A statistic evaluation on our test set revealed that the distribution of the minutiae quality values output 
by the MINDTCT is very uneven; 
values in certain ranges occur very frequently while others (e.g., between $0.5$ and $0.57$) are almost never assumed. 


The average number of minutiae detected in a single fingerprint depends on the sensor used, the feature extractor algorithms, the quality of the 
images, and even the finger type (e.g. thumbs contain more minutiae than other fingers). In our tests, we detected an average number of $84$ minutiae per fingerprint 
(excluding thumbs) inside ellipse $\mathcal{E}$.
Based on this number and the distribution of quality values, 
we can estimate the expected number $\tau$ of minutiae in a query fingerprint after filtering with 
minimum quality value $Q$. The results are listed in Table \ref{tab:qual1}.

 \begin{table}[tb]
\begin{center}
\caption{Expected number of minutiae in a query fingerprint after filtering with minimum quality value $Q$.}
\label{tab:qual1}

\begin{tabular}{|c|c|}
\hline
Minimum quality $Q$ & Av. no. $\tau$ of minutiae \\ \hline \hline
0   & 70 	\\
0.1 & 67	\\
0.2 & 52 	\\
0.3 & 48 	\\
0.4 & 41 	\\
0.5 & 33	\\
0.6 & 32 	\\  \hline
\end{tabular}
\end{center}
\end{table}

\begin{figure}[bt]                                  
\centering                                          
\includegraphics[width=\linewidth]{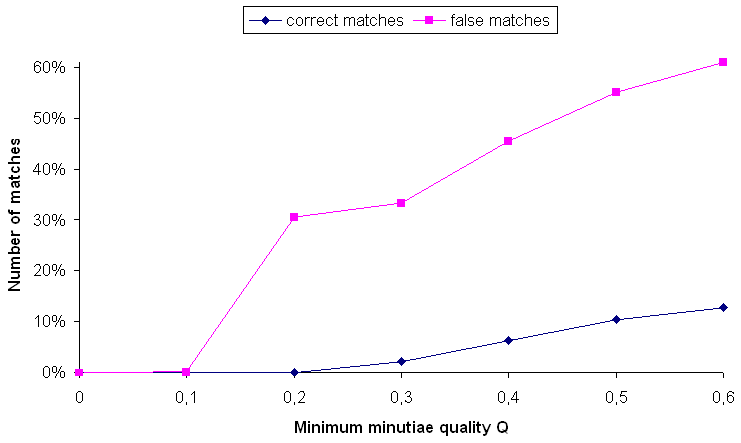}
\caption{Reduction of the correct and false matches by quality filtering for $f=6$, $t=120$, $r=400$ and 
$\delta_{\mathrm{v}}=7$.}            
\label{fig:qualgraph}
\end{figure}

The reduced average number $\tau$ of minutiae per query finger given to the matching algorithm 
results in a decreased average number $s$ of surplus minutiae per finger and in
less false matches (i.e., matches with chaff points). 
On the other hand, it may also reduce the number of correct matches (and likewise the match rate) because the minutiae filtered out could have matched with genuine minutiae in the vault. For different sets of parameters we
empirically determined the decrease of the number of correct 
and false matches resulting from the quality filtering. An example plot is presented in Figure \ref{fig:qualgraph}. 

For larger $r$ and smaller $\delta_{\mathrm{v}}$, quality filtering with higher values of $Q$ results in a more drastic reduction of 
the correct matches. Nevertheless, for various parameters we consistently found a value $Q$ between 0.2 and 0.3 to be optimal, reducing 
the false matches by approximately $30\%$ while decreasing the number of correct matches by less than $3\%$.

 




\subsection{Effect of minimum number of minutiae per finger}\label{res_chi}
\begin{figure}[bt]                                  
\centering                                          
\includegraphics[width=\linewidth]{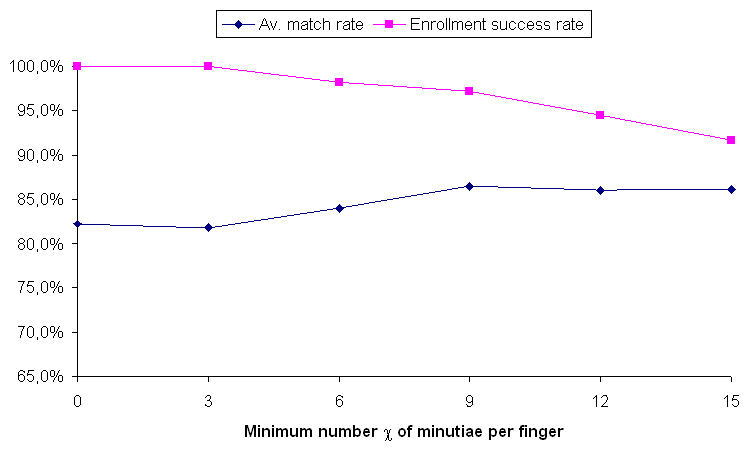}
\caption{Impact of enforcing a minimum number $\chi$ of minutiae per finger in $T$ to match rate and rate of successful enrollment for $f=6$, $u=4$, $t=120$, $r=400$, $\delta_{\mathrm{e}}=10$, $\delta_{\mathrm{v}}=7$ and $Q=0$.}            
\label{fig:chi}
\end{figure}

If the tolerance parameter $\delta_{\mathrm v}$ is set appropriately as described in Section \ref{delta}, the number of correct 
matches typically exceeds the number of false matches. 
On the other hand, if the matching algorithm fails to identify the correct isometry, the number of 
correct matches is typically significantly lower than the number of false matches. As explained in Section \ref{chi}, the enforcement of a 
minimum number $\chi$ of minutiae per finger in the template $T$ aims at reducing the frequency of such cases. We evaluated the effectiveness 
and reasonable configuration of this optimization by determining the ratio of fingers for which the number of false matches exceeded the number 
of correct matches for various values of the parameter $\chi$. 
Furthermore, we analyzed the influence of this optimization to the FTE by determining the rate at which a finger contained at least $\chi$ minutiae 
and, hence, would succeeded to enroll. 
The results of this evaluation are displayed in Figure \ref{fig:chi} by the curves of the match rate and the rate of successful enrollment. (Other failures of 
enrollment, particularly cases, where the fingers of a person contained less than $t$ minutiae in total, were neglected.)  Obviously, 
$\chi=9$ already yields a considerable improvement with only moderately increased FTE rates. 

\begin{figure}[bt]                                  
\centering                                          
\includegraphics[width=\linewidth]{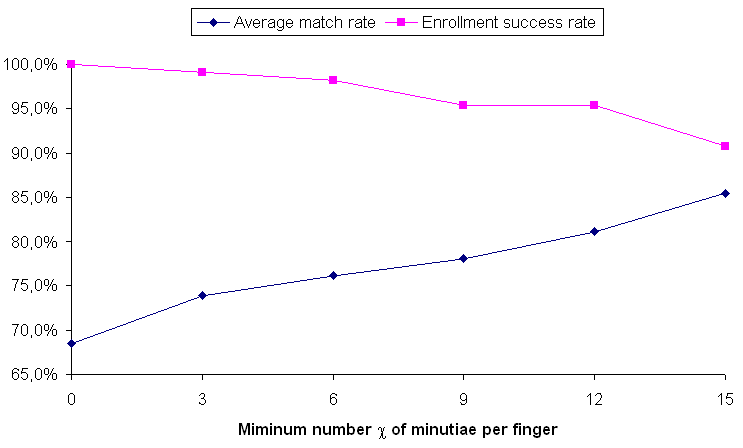}
\caption{Impact of enforcing a minimum number $\chi$ of minutiae per finger in $T$ to match rate and rate of successful enrollment for $f=6$, $u=4$, $t=100$, $r=600$, $\delta_{\mathrm{e}}=10$, $\delta_{\mathrm{v}}=7$ and $Q=0.3$.}            
\label{fig:chi2}
\end{figure}

The impact of the value of $\chi$ becomes particularly strong as the average number of false matches approaches the number 
of correct matches. As shown in Figure \ref{fig:chi2} for $f=6$, $u=4$, $t=100$, $r=600$, $\delta_{\mathrm{e}}=10$, 
$\delta_{\mathrm{v}}=7$ and $Q=0.3$, where even for  
$\chi=15$ the fraction between the average numbers of correct and false matches was 2.1 (as opposed to a fraction of 2.9 for the parameters of Figure \ref{fig:chi}), the average match rate steadily and considerably increases until $\chi=15$. The decrease of the successful enrollment rate 
is similar to the case of Figure \ref{fig:chi}. 
This finding indicates that in these cases it may be worth to choose $\chi$ larger than 9 at the cost of higher FTE rates. 

\subsection{Effect of pre-alignment of fingerprints}
We evaluated the impact of the pre-alignment of the fingerprints on our test set. We used two sets of parameters with 6 fingers ($f=6$), 
the first one being $f=6$, $u=4$, $t=120$, $r=400$, $\chi=15$ and $\delta_{\mathrm{v}}=7$ and the second one being 
$f=6$, $u=4$, $t=100$, $r=455$, $\chi=13$ and $\delta_{\mathrm{v}}=5$.

On average, the pre-alignment reduced the absolute rotation identified by the minutiae matching algorithm 
from $4.3^{\circ}$ to $2.7^{\circ}$. Much more important than that is the increasing correlation of a failure of the matching algorithm to 
correctly map the minutiae sets (resulting in more false than correct matches) 
to the size of the rotation angle. This allows setting up a threshold for the absolute rotation 
to reliably detect and reject individual query fingerprints for which the minutiae matching algorithm could not determine the correct mapping. 
In empirical tests, we evaluated the impact of the pre-alignment to the 
reliability at which a rotation threshold can sort out ``bad'' fingerprints that contribute more 
false than correct matches. Without the pre-alignment, a rotation threshold of $10^{\circ}$ rejected 
$86\%$ of those ``bad'' fingerprints, at the cost of $35\%$ false positives, i.e. ``good'' fingerprints being rejected. In contrast, when 
using the pre-alignment described in Section \ref{prealignment}, a rotation threshold of $8^{\circ}$ rejected $100\%$ of the ``bad'' fingerprints 
at the cost of only $20\%$ false positives. 

\subsection{Balancing correct and false matches}\label{delta}
In order to enable the minutiae matching algorithm to determine the correct isometry by which the query fingerprint is 
correctly aligned to the minutiae in the vault, we must ensure that, on average, the number of correct matches considerably 
exceeds the number of false matches. The results of Section \ref{res_chi} indicate that a fraction of 2 between the average numbers of correct and 
false matches already requires large values for $\chi$ which considerably increases the FTE.  

In \cite{MIKNS10}, the expected number $m_{\mathrm{f}}$ of false matches is estimated by 
$(r-t) s V_{\delta_{\mathrm v}}/|\mathcal{E}|$, where 
$V_{\delta}= 1+4 \sum_{i=1}^{\lceil \delta-1 \rceil}\left\lceil \sqrt{\delta^2-i^2} \right\rceil$
is the number of integer points in the 2-dimensional plane with Euclidean norm smaller than $\delta$ and $s$ is the average number of surplus minutiae 
(i.e., minutiae not matching with genuine minutiae) per query fingerprint. On the other hand, we can estimate 
$s \approx \tau -\mu t / f$, where $\tau$ is the average number of minutiae per query fingerprint after quality filtering. 

Our experiments show that for typical 
parameters the average number of false matches is 20\%-60\% larger than these estimations imply, depending on the specific parameters. The deviation 
is presumably due to those outliers resulting from an incorrect determination of the isometry: if the matching algorithm
is unable to detect the correct alignment, its optimization strategy with respect to the number of matches will yield extraordinary 
many false matches. Based on this observation, we adjust our above estimation to 
\begin{equation}
m_{\mathrm{f}} \approx 1.4 (r-t) (\tau- \mu t / f) V_{\delta_{\mathrm v}}/|\mathcal{E}|.
\label{eq:m_f}
\end{equation}  
Yet, we expect the number of false matches to grow linearly with $V_{\delta_{\mathrm v}}$, which is a quadratic 
function in $\delta_{\mathrm v}$.    

On the other hand, the average number $m_{\mathrm{c}}$ of correct matches is given by $\mu t$, where $\mu$ is the match rate, and therefore, 
grows slowly with increasing $\delta_{\mathrm v}$ as shown in Table \ref{tab:2}. Therefore, the selection of $\delta_{\mathrm v}$ should 
carefully balance 
the expected numbers of correct and false matches. For $f=3$, $u=4$, $t=66$, $r=320$ and $Q=0.3$, and for $5\leq \delta_{\mathrm v} \leq 15$ we 
estimated the number of false matches by (\ref{eq:m_f}) and the number of correct matches as $\mu t$ using the match rates 
empirically determined. The results show that, for these parameters, $\delta_{\mathrm v}\leq 8$ should be selected to ensure that the average number 
of correct matches is at least twice the number of correct matches.

\begin{figure}[bt]                                  
\centering                                          
\includegraphics[width=\linewidth]{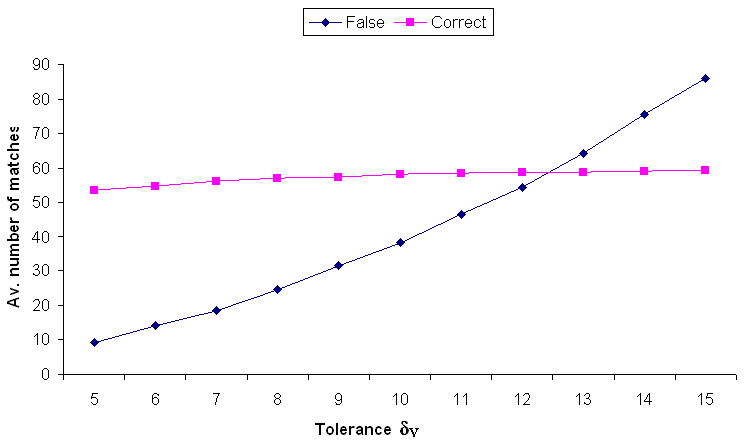}
\caption{Impact of the tolerance parameter $\delta_{\mathrm v}$ on the estimated average number of correct and false matches for $f=3$, $u=4$, $t=66$, $r=320$ and $Q=0,3$.}            
\label{fig:delta}
\end{figure}   

For a smaller ratio $r/t$, the curves meet at higher values of $\delta_{\mathrm v}$, but still the accelerating growth of the number of false 
matches implies that $\delta_{\mathrm v}\leq 8$ is a good choice.

\subsection{Provable security}
In \cite{MIKNS10}, minimum match rates are estimated that are necessary to provide provable security based on the lower bounds given in 
\cite{DORS08}. These minimum match rates depend on the tolerance parameter $\delta_{\mathrm{v}}$ and the number $s$ of surplus minutiae in the query 
fingerprint (i.e., the minutiae not matching with genuine minutiae of the template) by which the number of false matches is estimated. 
A comparison of these minimum match rates with the empirically 
determined rates in  Table \ref{tab:2} (which have been roughly confirmed in the presence of chaff points with appropriate optimizations applied) shows 
that in practice a provable secure fuzzy fingerprint vault could only be possible for $\delta_{\mathrm{v}}\leq 7$ and $s \leq 20$. 
As argued in Section \ref{delta} we can estimate $s \approx \tau-\mu t / f$, where $\tau$ is the number of minutiae of the query fingerprint after quality filtering. Furthermore, the results of Section \ref{res_qual} show that quality filtering should not be 
expected to reduce $\tau$ significantly below 48. This implies that $s=20$ can only be achieved for $\mu t/f \geq 28$. However, evaluation of 
the values $t/f$ and $\mu$ given in Tables \ref{tab:1} and \ref{tab:2}, respectively, for $\delta_{\mathrm{v}}\leq 7$ reveals that this condition 
is only met for $u\leq 2$, where the match rates are much lower than the minimum match rates derived in \cite{MIKNS10}. 

Furthermore,  as argued in Section \ref{delta}, the false match rates observed in practice are 20\% to 60\% higher than the estimation 
in \cite{MIKNS10} would imply. 
This means that the value for $s$ must even be  decreased to 12-17 to obtain provable security with the minimum match rates given for $s=20$. 

As a summary, we conclude that provable security seems out of reach, unless the average number of surplus minutiae per query fingerprint can be 
further reduced by improved quality filtering methods. 

\subsection{Selection of parameters}\label{parameters}
For the selection of eligible parameters for a fuzzy fingerprint vault secure against existing attacks, we suggest the following method:
\begin{enumerate}
\item Define the number of fingers used.  
\item Set the number of captures per finger for enrollment to $u=3$ or, preferably, $u=4$. 
\item Choose $\delta_{\mathrm e}$ between 5 and 15, and $t\leq f M_r(\delta_{\mathrm e})$, where $M_r(\delta_{\mathrm e})$ is the 
 number of reliable minutiae for $\delta_{\mathrm e}$ indicated in Table \ref{tab:1}. The choice of $\delta_{\mathrm e}$ and $t$ should be carefully tested with respect to the FTE rate. 
\item Choose $\delta_{\mathrm v}$ between $5$ and $7$. Smaller values may drastically reduce the match rates, which makes 
it difficult to achieve a high security level. (Note, that setting $\delta_{\mathrm v}$ much smaller than $\delta_{\mathrm e}$ may result in smaller match rates than indicated in Table \ref{tab:2}.) Larger values will result in too many false matches (see Section \ref{delta}).   
\item Set the minimum distance $d$ between the minutiae and chaff points in the template to approximately $(3/2) \delta_{\mathrm v}$ to reduce 
the probability that a minutiae in the query fingerprint is closer to a chaff point than to its counterpart in $T$.
\item Set $Q$ between $0.2$ and $0.3$.
\item For a broad range of values for $r$, numerically estimate the expected numbers $m_{\mathrm{c}}$ and $m_{\mathrm{f}}$ of correct and 
false matches, respectively, by $\mu t$ and (\ref{eq:m_f}) using
 match rate $\mu$ indicated for the selected $u$ and $\delta_{\mathrm v}$ in Table \ref{tab:2} and the estimate $\tau \approx 50$. 
 Then, for each value of $r$ set $k$ 10\%-25\% smaller than $m_{\mathrm{c}}- m_{\mathrm{f}}$ (see Section \ref{match_rate}) and compute the security against existing attacks as 
 $129 \zeta(t,\chi) k \log^2{(k)} (r/t)^k$. Select the pair $r, k$ for which the security is maximized.
\item Select $\chi=9$, if the fraction between the estimates for the average number of correct and false matches is at least 2.7. Otherwise, increase 
$\chi$ up to 15, depending on this  fraction. However, ensure that the fraction is at least 2, if necessary, by decreasing $r$ or $\delta_{\mathrm v}$.
\item If the maximum security determined is higher than the level aimed at, lower $r$ and $k$ to reduce the false rejection rate (FRR), and decrease $t$ to reduce the failure to enroll (FTE) rate. 
\item Empirically evaluate the average number of correct and false matches. If these numbers significantly deviate from the estimations used in the 
previous step, repeat this step with appropriate correction factors.    
\end{enumerate}
 
Based on our statistical data and the method described above, the example parameters listed in Table \ref{tab:par} 
have been determined to provide the indicated 
security level against existing attacks. 
We did not experimentally determine real error rates during enrollment and verification; therefore, these 
parameters are mere suggestions which require practical validation. We set $d=\lfloor 3/2\cdot \delta_{\mathrm v}\rfloor$, $Q=0.3$ and, as the fraction computed in step 8 above was always greater than 3, $\chi=9$. Furthermore, we choose $r< \lfloor 0.2 \cdot 87000 / V_{d}\rfloor$, which is less than half of the maximum value possible, to avoid the attack described in \cite{CST06} (see Section \ref{practical}) that could significantly reduce security.

\begin{table}[tb]
\begin{center}
\caption{Parameters for a security level of $2^{Sec}$.}
\label{tab:par}
\begin{tabular}{|ccccccc|c|}
\hline
$f$ & $u$ &  $\delta_{\mathrm e}$ & $\delta_{\mathrm v}$ & $t$  & $r$ & $k$ & $Sec$ \\ \hline \hline
2 & 2 & 7 & 5 & 62 & 240 & 27 & 68\\
3 & 2 & 7 & 5 & 90 & 202 & 45 & 69\\
3 & 2 & 7 & 5 & 90 & 351 & 41 & 97\\
3 & 3 & 7 & 5 & 70 & 360 & 34 & 97\\ \hline
\end{tabular}
\end{center}
\end{table}

\subsection{Practical evaluation}
In order to test the performance of our fuzzy fingerprint vault and its optimizations under realistic circumstances, we enrolled $10$ persons 
using $6$ fingers 
(i.e., $f=6$) using the parameters $u=4$, $t=120$, $r=400$, $k=53$, $\delta_{\mathrm{e}}=10$, 
$\delta_{\mathrm{v}}=7$, $\chi=15$ and $Q=0.3$. For these parameters, the best known attack requires $2^{100}$ operations; thus, a security equivalent 
to $100$ bit keys is achieved. For the persons that had been successfully enrolled, we performed a verification. As capture device, we used an 
optical multi-finger sensor (Cross Match L SCAN Guardian), the NBIS package of NIST (in particular, NFSEG and MINDTCT) \cite{WGT07} for fingerprint segmentation and feature extraction. 

First, we evaluated the performance of the enrollment by its FTE and duration. It was aborted after $3$ unsuccessful attempts per finger and 
the FTE was computed as the fraction of aborted 
enrollments. Furthermore, we determined the average number of retries (repetition of capturing) needed per user (summed up for all fingers) and the 
time needed. 

In order to assess the reliability of the verification, we evaluated the FRR (without any repeated attempts). The FAR for this small sample was zero. 

We performed the simulation for two versions of the scheme independently: the first 
simulation was performed for a basic version without the quality filtering of the minutiae during 
verification (Section \ref{quality_filter}), the pre-alignment of fingerprints (Section 
\ref{prealignment}), and the enforcement of a minimum number of minutiae per finger in the 
template (Section \ref{chi}). 
The second version implemented these optimizations. The results of the simulation are listed in Table \ref{tab:sim}. 
\stepcounter{footnote}           
\footnotetext{The only failed verification succeeded in the second attempt.}  

\begin{table}[tb]
\begin{center}
\caption{Performance of the scheme in a simulation with 10 persons. Duration is given in minutes.}
\label{tab:sim}
\begin{tabular}{|c|ccc|c|}
\hline
& \multicolumn{3}{c|}{ Enrollment}  & Verification\\ 
 & FTE & Retries & Duration & FRR \\ \hline \hline
Basic   		& 20\% 	&	0.5 & 2.1  & $25\%$	 \\ 
Optimized 	& 0\% 	&	2 	& 3.5  	& $10\%^\thefootnote$\\ \hline
\end{tabular}
\end{center}
\end{table}

\section{Conclusions}\label{conclusions}

Our analysis shows that a fuzzy vault for multiple fingerprints can be very secure against template recovery from the helper data, if appropriate optimizations are applied. 
Filtering minutiae for reliability during enrollment and for quality during verification turn out to be particularly effective. Furthermore, 
enforcing a minimum number of minutiae per finger in the template significantly increases matching performance. Both optimizations are very 
sensitive to the respective thresholds, which must be carefully set on the basis of empirical data. Interestingly, we were able to solve the 
fingerprint alignment problem using a simple algorithm and without storing additional helper data or using singular point detection.  

Although we did not achieve match rates required to prove the security by information theoretic arguments as discussed in \cite{MIKNS10}, 
we can obtain a security level against existing attacks of $2^{70}$ for two fingers and of $2^{97}$ for three fingers. 

Our simulation of enrollment and verification indicates that the scheme can be effective and efficient in practice. The process of capturing several fingers can be facilitated 
using multi-finger sensors. Nevertheless, the parameters need to be selected with care to reduce the error rates and effort for enrollment. Furthermore,
simulation tests on a larger scale would be needed to obtain more reliable data on achievable error rates. In particular, it would be interesting 
to investigate the fraction of fingers or persons that consistently fail to provide a sufficient number of reliable minutiae.  

Finally, we would like to stress that our security analysis only covered template recovery attacks. 
Other types of attacks have been published \cite{sb07} and need to be addressed 
before the scheme can be considered ready for use. We encourage research on methods to harden the fuzzy fingerprint vault against 
these attacks.

\section*{Acknowledgments}
This work was conducted as part of the projects ``BioKeyS-Multi'' and ``BioKeyS Pilot-DB'' of the Bundesamt f\"ur Sicherheit in der Informationstechnik.

The matching algorithm was designed and implemented by Stefan Sch\"urmans. 

\bibliographystyle{IEEEtran}
\bibliography{IEEEabrv,Bibliography}

\end{document}